\documentstyle[aps]{revtex}
\begin {document}
\preprint{MKPH-T-98-7}
\title{
{\bf On Complete Sets of Polarization Observables}
\footnote{Supported by the Deutsche 
Forschungsgemeinschaft (SFB 201) and the National Science and 
Engineering Research Council of Canada}}

\author{Hartmuth Arenh\"ovel$^{1}$, Winfried Leidemann$^{2}$,
and Edward L. Tomusiak$^{3}$}
\address{
$^{1}$Institut f\"ur Kernphysik,
Johannes Gutenberg-Universit\"at,
 D-55099 Mainz, Germany\\
$^{2}$Dipartimento di Fisica, Universit\`a degli Studi di Trento, and\\
Istituto Nazionale di Fisica Nucleare, Gruppo collegato di Trento,
 I-38050 Povo, Italy\\
$^{3}$Department of Physics and Engineering Physics
and Saskatchewan Accelerator Laboratory\\
University of Saskatchewan, Saskatoon, Canada S7N 0W0
}
\maketitle
\begin{abstract}
A new criterion is developed which provides a check as to whether a chosen 
set of polarization observables is complete with respect to the determination 
of all independent $T$-matrix elements of a reaction of the type 
$a+b\rightarrow c+d+\cdots$. As an illustrative example, this criterion is
applied to the longitudinal observables of deuteron electrodisintegration. 
\end{abstract}
\pacs{PACS numbers: 13.88.+e, 24.70.+s}

\section{Introduction}
The major reason for studying polarization phenomena in reactions of the 
type $a+b\rightarrow c+d+\cdots\,$ lies in the fact that only the use of 
polarization degrees of freedom allows one to obtain complete information on 
all possible reaction matrix elements. Without polarization, the cross 
section is given by the incoherent sum of squares of the 
reaction matrix elements 
only. Thus small amplitudes are masked by the dominant 
ones. On the other hand, small amplitudes very often contain interesting 
information on subtle dynamical effects. This is the place where 
polarization observables enter, because such observables in general contain 
interference terms of the various matrix elements in different ways. 
Thus a small amplitude may be considerably amplified by the interference
with a dominant matrix element. An example is provided by the influence of
the small electric form factor of the neutron on the transverse in-plane 
component of the neutron polarization in quasi-free deuteron 
electrodisintegration \cite{ArC81,ArL88,Kle96}. 
It is just this feature for which polarization 
physics has become such an important topic in various branches of physics. 

In view of the great number of possible polarization observables, i.e., 
beam and target asymmetries and/or polarization components of various 
outgoing particles, it is natural to ask which set of observables allows 
in principle a complete determination of all reaction matrix elements. 
Such sets are called complete sets of observables for obvious reasons. 
The usual strategies follow the explicit construction of a particular 
complete set. But this is practical only for a small number of matrix 
elements as in pion photoproduction \cite{BaD75,KeW96}. 
With an increasing number of matrix elements as in deuteron photo- and 
electrodisintegration \cite{DmG89,ArS90,ArL93}, for example, this method 
becomes more and more complicated. A general necessity statement as to 
the selection of polarization observables has been proven by Simonius 
\cite{Sim67}, which excludes certain subsets from being complete. However, 
this statement is not specific enough. For this reason we want to study the 
alternative question: Given an appropriate subset of all observables, 
is there a criterion which allows one to decide whether this set constitutes 
a complete set of observables. 

First we consider in Sect.\ 2 a simple example. The essential idea is then
developed in Sect.\ 3 for the case of a general $n$-dimensional real 
quadratic form. In Sect.\ 4 we apply this criterion to the
observables of a general reaction $a+b\rightarrow c+d+\cdots\,$ by 
rewriting the hermitean forms of the observables into real quadratic forms
in the real and imaginary parts of the $T$-matrix elements. An illustrative 
example is considered for the longitudinal observables of deuteron 
electrodisintegration in Sect.\ 5. 

\section{A simple case for illustration}

We will begin by considering a reaction $a+b\rightarrow c+d+\cdots\,$ of 
particles having 
spins $s_a,\, s_b,\, s_c,\dots\,$ where angular momentum conservation requires 
that both $s_a+s_b$ and $s_c+s_d+\cdots\,$ are either integer or half-integer. 
Then the number of $T$-matrix elements for this reaction is given 
by $N=N(s_a)N(s_b)N(s_c)\cdots\,$ where for massive particles one has 
\begin{eqnarray}
N(s_\alpha)=2s_\alpha+1\,,
\end{eqnarray}
while for massless particles
\begin{eqnarray}
N(s_\alpha)=2\,.
\end{eqnarray}
In case parity conservation holds, the number of independent $T$-matrix 
elements reduces to $n=N/2$ or $n=(N-1)/2$, depending on whether $N$ is even 
or odd. Since any observable is given as a hermitean form in the $T$-matrix 
elements, the number of linearly independent observables is $n^2$. This does, 
however, not mean, that these observables are independent of each other in a 
more general sense, because on can find quadratic relations between them.
In fact, since each $T$-matrix element is in general a complex number, but one 
overall phase is undetermined, one has the complete information on the 
reaction contained in a set of $2n-1$ real numbers. It is therefore argued 
that a set of $2n-1$ properly chosen observables should suffice 
to determine completely all 
$T$-matrix elements. However, the solution is in general not unique but 
contains discrete ambiguities due to the fact that the solution implies 
the determination of the roots of a higher order polynomial, and therefore 
one needs additional information from further observables, like relative 
magnitudes etc.\ \cite{BaD75,KeW96}. 

We will illustrate this feature by considering the simplest nontrivial 
case of two independent complex matrix elements $T_1$ and $T_2$. 
It can be realized by the absorption of a scalar particle by a $s=1/2$ 
particle leading to a 
$s=1/2$ particle. In this case any observable is represented by a $2\times 2$ 
matrix for which we may choose besides the unit matrix $(\sigma_0)$ the 
three Pauli matrices $(\sigma_i,\, i=1,2,3)$. Thus we have a set of four 
linearly independent observables 
\begin{eqnarray}
{\cal O}_\alpha=T^\dagger \sigma_\alpha T\,,\quad (\alpha=0,\dots,3)\,,
\end{eqnarray}
or in greater detail
\begin{eqnarray}
{\cal O}_0&=&|T_1|^2+|T_2|^2\,,\\
{\cal O}_1&=&2\,\Re e(T_1^*T_2)\,,\\
{\cal O}_2&=&2\,\Im m(T_1^*T_2)\,,\\
{\cal O}_3&=&|T_1|^2-|T_2|^2\,.
\end{eqnarray}
Evidently, the following relations hold 
\begin{eqnarray}
|T_1|^2|T_2|^2&=& \frac{1}{4}({\cal O}_1^2+{\cal O}_2^2)\,,\\
{\cal O}_3^2 &=& {\cal O}_0^2 - 4|T_1|^2|T_2|^2\nonumber\\
             &=& {\cal O}_0^2 - ({\cal O}_1^2+{\cal O}_2^2)\,,
\label{relation2}
\end{eqnarray}
and therefore
\begin{eqnarray}
{\cal O}_0^2&\ge&{\cal O}_1^2+{\cal O}_2^2\,.\label{relation1}
\end{eqnarray}
The relation in (\ref{relation2}) constitutes just the above mentioned 
quadratic relation between observables for this case. 

Without loss of generality we can choose the undetermined 
overall phase such that $T_1$ becomes real and is $>0$. Then one can 
determine $T_1$ and $T_2$ from three 
observables, for example from ${\cal O}_0$, ${\cal O}_1$ and ${\cal O}_2$. 
Explicitly one finds 
\begin{eqnarray}
T_1^2&=&\frac{1}{2}\Big({\cal O}_0\pm\sqrt{{\cal O}_0^2 -({\cal O}_1^2+
        {\cal O}_2^2)}\,\Big)\,,\label{eqt1}\\
T_2 &=& \frac{1}{2T_1}({\cal O}_1+i{\cal O}_2)\,.
\end{eqnarray}
Because of the relation (\ref{relation1}), the right hand side of (\ref{eqt1})
is nonnegative and, therefore, the quadratic equation (\ref{eqt1}) for 
$T_1$  yields in general two real solutions. This is the discrete ambiguity 
mentioned in the introduction. In order to determine the proper solution, 
one has to consider the observable ${\cal O}_3$. Inserting the general 
solution, one finds
\begin{eqnarray}
{\cal O}_3&=&\pm\sqrt{{\cal O}_0^2 -({\cal O}_1^2+{\cal O}_2^2)}
\,,
\end{eqnarray}
which corresponds to the relation (\ref{relation2}). 
Thus the sign of ${\cal O}_3$ selects in this case the proper solution. 

An alternative way to determine the $T$-matrix elements from the observables 
is to express all bilinear products of the form $T_i^*T_j$ by linear 
expressions in the observables \cite{ArS90}, which reads in the considered case
\begin{eqnarray}
T_1^*T_1 &=& \frac{1}{2}({\cal O}_0+{\cal O}_3)\,,\\
T_1^*T_2 &=& \frac{1}{2}({\cal O}_1+i{\cal O}_2)\,,\\
T_2^*T_2 &=& \frac{1}{2}({\cal O}_0-{\cal O}_3)\,.
\end{eqnarray}
Assuming again $T_1$ real and $>0$, one finds as solution
\begin{eqnarray}
T_1&=&\sqrt{\frac{1}{2}({\cal O}_0+{\cal O}_3)}\,,\\
T_2&=&\frac{{\cal O}_1+i{\cal O}_2}{\sqrt{2({\cal O}_0+{\cal O}_3)}}\,,
\end{eqnarray}
which coincides with the previous solution. Evidently, the quadratic 
relation between the observables is fulfilled by this explicit solution. 
In this case, one obtains directly the explicit form of the $T$-matrix elements
expressed in terms of observables. The disadvantage of this approach is 
that one has no choice with respect to the observables to be measured 
because they are determined by the linear expressions of $T_i^*T_j$. 

\section{General criterion for a complete set of real quadratic forms}
In this section we will study the following question: Given an $n$-dimensional
real vector $x=(x_1,\dots,x_n)$ and $n$ real quadratic forms 
\begin{eqnarray}
f^\alpha(x) &=& \frac{1}{2}\sum_{ij}x_i H^\alpha_{ij} x_j\,,
\end{eqnarray}
where $\alpha=1,\dots,n$, what is the criterion that the set of $n$ equations 
\begin{eqnarray}
f^\alpha(x) &=& c^\alpha\label{quadrel}
\end{eqnarray}
for given constants $c^\alpha$ possesses a solution $x^0$ with allowance for 
possible discrete ambiguities? In other words, is the set of equations 
in (\ref{quadrel}) invertible? A sufficient condition for the inversion is 
that  in the vicinity of $x^0$ the Jacobian $Df|_{x^0}$ is nonvanishing, 
i.e. 
\begin{eqnarray}
Df|_{x^0}&=& \left|\frac{\partial f^\alpha}{\partial x_j}\right|\neq 0\,.
\end{eqnarray}
This means that the $n$ vectors $v^\alpha$ defined 
by the partial derivatives 
\begin{eqnarray}
\frac{\partial f^\alpha}{\partial x_j}&=&H^\alpha_{ji}x^0_i\,,
\end{eqnarray}
according to 
\begin{eqnarray}
v^\alpha=\left( \begin{array}{c} \frac{\partial f^\alpha}{\partial x_1}
\\ \vdots \\ \frac{\partial f^\alpha}{\partial x_n} \end{array}\right)
=\left( \begin{array}{c} H^\alpha_{1i}x^0_i \\ \vdots \\ 
H^\alpha_{ni}x^0_i \end{array}\right)\,,
\end{eqnarray}
are linearly independent. The question thus is: under which condition is this 
true for an arbitrary $x^0$? The $n$ vectors $v^\alpha$ are linear
forms in $x^0$
\begin{eqnarray}
v^\alpha=
\left(\begin{array}{c} H^\alpha_{11} \\ \vdots \\ H^\alpha_{n1}
\end{array}\right)x^0_1
+\cdots +
\left(\begin{array}{c} H^\alpha_{1n} \\ \vdots \\ H^\alpha_{nn}
\end{array}\right)x^0_n
\,,
\end{eqnarray}
which we may write in the form
\begin{eqnarray}
v^\alpha=\sum_{k=1}^nw^{\alpha}(k) x^0_k\,, \label{valpha} 
\end{eqnarray}
where we have introduced
\begin{eqnarray}
w^{\alpha}(k)=
\left(\begin{array}{c} H^\alpha_{1k} \\ \vdots \\ H^\alpha_{nk}
\end{array}\right)
\,.
\end{eqnarray}
A necessary and sufficient condition for the linear independence of the $n$ 
vectors $v^\alpha$ is that the determinant built from the $n$ vectors 
$v^\alpha$ is nonvanishing
\begin{eqnarray}
\det\,[v^1,\dots,v^n]\neq 0\,,
\end{eqnarray}
which with the help of (\ref{valpha}) is written in the equivalent form
\begin{eqnarray}
\sum_{k_1=1}^n\cdots\sum_{k_n=1}^n x^0_{k_1}\cdots x^0_{k_n} 
\det\,[w^1(k_1),\dots,w^n(k_n)]\neq 0\,.
\end{eqnarray}
Evidently, a necessary condition for the nonvanishing of this determinant
is that of the various sets of $n$ vectors 
\begin{eqnarray}
W(k_1,\dots,k_n)=\{w^{\alpha}(k_\alpha);\,\alpha=1,\dots,n;\,k_\alpha\in
\{1,\dots,n\}\}
\end{eqnarray}
at least one is a set of linearly independent vectors, i.e.,
\begin{eqnarray}
\det\,[w^1(k_1),\dots,w^n(k_n)]=
\left|\begin{array}{ccc} H^1_{1k_1} & \cdots & H^n_{1k_n} \\ 
\vdots & &\vdots \\ H^1_{nk_1} & \cdots & H^n_{nk_n} \end{array}\right|\neq 0
\,.
\end{eqnarray}
Note that the $k_\alpha$ need not be different. If none of these 
determinants is nonvanishing, the $n$ vectors $v^{\alpha}$ certainly 
cannot be linearly independent. 
Moreover, if this condition is fulfilled exactly for one set of 
$\{k_\alpha\}$ while the determinants of all other $W(k_1,\dots,k_n)$ 
vanish, then this condition is also sufficient. In case one finds 
$|W(k_1,\dots,k_n)|\neq 0$ for more than one set $\{k_\alpha\}$, then 
it could in principle happen that for a specific choice 
of $x^0$ the set $v^\alpha$ becomes linearly dependent, although it will 
in general be rather unlikely to happen. But one has to check this in a 
given situation. 
This is the criterion we were looking for and which now we will apply to the 
hermitean forms of a set of observables. 

\section{Complete sets of observables}
We represent an observable by an $n\times n$ hermitean form $f^\alpha$ 
in the complex $n$-dimensional variable $z$  
\begin{eqnarray}
f^\alpha(z)&=&\frac{1}{2}\sum_{j'j} z_{j'}^* H_{j'j}^\alpha z_j\,,
\end{eqnarray}
where hermiticity requires 
\begin{eqnarray}
(H_{j'j}^{\alpha})^*=H_{jj'}^\alpha\,.
\end{eqnarray}
Now we will rewrite this form into a real quadratic form by introducing
\begin{eqnarray}
z&=&x+i y\,,\\
H^\alpha&=&A^\alpha+i\, B^\alpha \,,
\end{eqnarray}
where $A^\alpha$ and $B^\alpha$ are real matrices and $A^\alpha$ is 
symmetric whereas $B^\alpha$ is antisymmetric.
Considering further the fact that one overall phase is arbitrary, we may 
choose $y_{j_0}=0$ for one index $j_0$ and then obtain
\begin{eqnarray}
f^\alpha(x+iy)&=&\frac{1}{2}\Big[\sum_{j'j} x_{j'} A_{j'j}^\alpha x_j
+\sum_{\tilde j'\tilde j} y_{j'} A_{j'j}^\alpha y_j 
+2\sum_{\tilde j'j} y_{j'} B_{j'j}^\alpha x_j\Big]
\,,
\end{eqnarray}
where the tilde over a summation index indicates that the index $j_0$ has to  
be left out. Introducing now an $m$-dimensional real vector $u$ $(m=2n-1)$ by 
\begin{eqnarray}
u=( x_1, \dots , x_n , y_1, \dots , y_{j_0-1}, y_{j_0+1}, \dots , y_n)\,,
\end{eqnarray}
one can represent the $n\times n$ hermitean form by an $m\times m$
real quadratic form
\begin{eqnarray}
\widetilde f^\alpha(u)&=&\frac{1}{2}\sum_{l'l=1}^{m} u_{l'} 
                     \widetilde H_{l'l}^\alpha u_l\,,
\end{eqnarray}
where we have further defined the $m\times m$ matrix 
\begin{eqnarray}
\widetilde H^\alpha=\left(\begin{array}{cc} A^\alpha & 
(\widetilde B^\alpha)^T \\
\widetilde B^\alpha &  {\hat A}^\alpha \end{array}\right)\,.
\end{eqnarray}
Here $\widetilde B^\alpha$ is obtained from $B^\alpha$ by cancelling the 
$j_0$-th row, and ${\hat A}^\alpha$ from $A^\alpha$ by cancelling 
the $j_0$-th row and column. Thus $\widetilde B^\alpha$ is an $(n-1)\times n$ 
matrix and $ {\hat A}^\alpha$ an $(n-1)\times(n-1)$ matrix. 

Application of the above criterion means now the following. For a given set of 
$m$ observables one first has to construct the $m\times m$ 
matrices $\widetilde H^\alpha$, 
and then one builds from their columns according to the sets 
$\{k_1,\dots,k_m\}$ the matrices  
\begin{eqnarray}
\widetilde W(k_1,\dots,k_m)=\left(\begin{array}{ccc} 
\widetilde H^1_{1k_1} & \cdots &\widetilde  H^m_{1k_m} \\ 
\vdots & &\vdots \\ \widetilde H^1_{mk_1} & \cdots & \widetilde H^m_{mk_m} 
\end{array}\right)
\,.
\end{eqnarray}
If at least one of the determinants of 
$\widetilde W(k_1,\dots,k_m)$ is 
nonvanishing then one has a complete set, up to the 
already mentioned discrete ambiguities. 

We will illustrate this for the above considered simplest nontrivial case
$n=2$ choosing ${\cal O}_0$, ${\cal O}_1$, and ${\cal O}_2$ for which one 
has with $j_0=1$
\begin{eqnarray}
\widetilde \sigma_0=\left(\begin{array}{ccc} 1 & 0 & 0 \\ 0 & 1 & 0 \\
0 & 0 & 1 
\end{array}\right)
\,,\quad 
\widetilde \sigma_1=\left(\begin{array}{ccc} 0 & 1 & 0 \\ 1 & 0 & 0 \\
0 & 0 & 0 
\end{array}\right)
\,,\quad 
\widetilde \sigma_2=\left(\begin{array}{ccc} 0 & 0 & 1 \\ 0 & 0 & 0 \\
1 & 0 & 0 
\end{array}\right)\,.
\end{eqnarray}
One finds in this case three $\widetilde W$ matrices which possess 
a nonvanishing determinant 
\begin{eqnarray}
\widetilde W(1,1,1)=\left(\begin{array}{ccc} 1 & 0 & 0 \\ 0 & 1 & 0 \\
0 & 0 & 1 
\end{array}\right)\!, 
\widetilde W(2,2,1)=\left(\begin{array}{ccc} 0 & 1 & 0 \\ 1 & 0 & 0 \\
0 & 0 & 1 
\end{array}\right)\!,
\widetilde W(3,1,3)=\left(\begin{array}{ccc} 0 & 0 & 1 \\ 0 & 1 & 0 \\
1 & 0 & 0 
\end{array}\right)\!.
\end{eqnarray}
Thus, the three observables constitute a complete set. For this case, any group
of three observables is complete and the direct solution is simpler. The 
usefulness of this new criterion becomes apparent if one considers reactions
with a larger number of independent $T$-matrix elements, as is done in the next
section. 

\section{Application to the longitudinal observables in electromagnetic 
deuteron break up}

Now we will consider the more complex case of deuteron 
electrodisintegration. An extensive formal discussion of polarization 
observables may be found in \cite{DmG89,ArL93}. The total number of $T$-matrix 
elements is $3\times 3\times 4=36$ which is reduced by parity conservation to 
$n=18$. Of these, 6 are associated with the charge or longitudinal current 
density component while the remaining 12 belong to the transverse current 
density components. Only the latter appear in photodisintegration. In order 
to simplify the present discussion, we will restrict ourselves to 
the purely longitudinal contribution. 

We would like to remind the reader that the differential cross section and 
the polarization components of the outgoing particles are completely 
described by the so-called structure functions, which can be separated 
experimentally by appropriate kinematic settings as is discussed in detail 
in \cite{ArL95}. For the present discussion, only the longitudinal 
ones will be considered. In the following, we will use the term 
``observable'' for a longitudinal structure function
$f_L^{(')IM} (X)$, where $X=1$ refers to the differential cross section, 
$X=x_j$, $y_j$, and $z_j$ to the respective polarization components 
$P_{x}(j)$, $P_{y}(j)$, and $P_{z}(j)$ of one outgoing nucleon
$(j=1,2)$, and $X=x_ix_j$ to the polarization components $P_{x_i}P_{x_j}$ 
of both outgoing nucleons. Since one has six independent longitudinal 
$T$-matrix elements, each observable is represented by a hermitean 
$6\times 6$ matrix whose explicit form depends on the adopted basis 
for the $T$-matrix and the chosen reference frame. 

We will use here the same definition of the polarization components as 
in \cite{ArL93} which is based on the Madison convention. Our standard 
form of the $T$-matrix is characterized by the 
quantum numbers $\{s,m_s,\lambda,m_d\}$ of the total spin of the two outgoing 
nucleons, its projection on their relative momentum, the spin projections 
of the virtual photon and the deuteron on the momentum transfer, respectively. 
With respect to  the definition of the observables 
in terms of the $T$-matrix elements, we refer to \cite{ArL93} where a 
complete listing of all possible observables is given. 
However, for the present discussion the representation of the $T$-matrix 
in the helicity basis is more advantageous, because the hermitean matrices 
of the observables have a simpler structure. In this case, the 
quantum numbers of the $T$-matrix elements are given by the spin 
projections of the participating particles on the respective momenta 
$\{\lambda_p,\lambda_n,\lambda,\lambda_d\}$.
We have numbered the independent matrix elements according to the listing in 
Table \ref{qnlist}. 

For the longitudinal observables, one finds a total number of 72 whereas 
the total number of linearly independent observables is 36. It is obvious that 
the observables of a complete set should be selected from linearly 
independent ones. A possible choice 
for a set of linearly independent observables is listed in Table \ref{tab1}, 
and we will restrict the following discussion to this set. However, any other 
set of linearly independent observables, as may be obtained from the total 
number of 72 by taking into account the linear relations of \cite{ArL93}, 
can also be used. For the observables of type $A$ in Table \ref{tab1}, 
one has five $IM$-values and for type $B$ four (see \cite{ArL93}).

Now a complete set of longitudinal observables in the sense 
described in the introduction should contain 11 observables only, but 
the question is which ones. In order to select a proper subset of 11 
from a set of 36 linearly independent observables, we have now applied our new 
criterion by first constructing the corresponding matrices 
$\widetilde H^\alpha$ and then by checking the determinants of 
the matrices $\widetilde W(k_1,\dots,k_{11})$ for various choices 
of $(k_1,\dots,k_{11})$. First, we did a restricted 
search by setting $j_0=1$ and $k_\alpha=1,\, (\alpha=1,\dots,11)$, and 
have found as possible complete sets of observables those 
listed in Table \ref{comset}. Here and in the following tables the 
superscripts refer to the $IM$-values of the corresponding observables.
The meaning of this table is that 
from each of the 11 columns one may pick arbitrarily one of the listed 
observables in order to form a complete set. Other possible complete sets, 
which are obtained by setting $j_0=3$ and $k_\alpha=3,\, (\alpha=1,\dots,11)$,
are listed in Table \ref{comseta}. Finally, allowing for more general 
choices for $(k_1,\dots,k_{11})$, we found that
there is only a very weak restriction on the possible
sets. In fact, in order to obtain a complete set one may select from the total
36 observables almost any set of 11 observables. The only limitation is
that the set should not include more than eight of the 12 observables
which are listed in Table \ref{rstrctset}, since otherwise the set would 
be incomplete. This restriction is a consequence of a specific structure 
of the matrices $A^\alpha$ and $B^\alpha$ for the listed observables. 
We note in passing, that all these complete sets fulfil the necessity
criterion of Simonius \cite{Sim67}.

Any complete set of 11 observables leads to a system of 11 nonlinear
coupled equations for the $T$-matrix elements. 
For the sets listed in Table \ref{solve} 
we solved the resulting equations numerically by using as input the 
observables resulting from a theoretical calculation as presented in 
\cite{ArL92,ArL95}, where we have chosen arbitrarily a specific internal 
excitation energy ($E_{np}=100$ MeV) and momentum transfer 
($q_{c.m.}^2=5$ fm$^{-2}$) for various n-p angles $\theta_{np}$. 
The first two sets are cases from Tables \ref{comset} and \ref{comseta}, 
while the two last sets do not correspond to the application of our criterion 
with a specific choice of a column $k_{\alpha}=k_0$. In general, we found 
a larger number of solutions which is a manifestation of the previously 
mentioned discrete ambiguities. On the other hand, we already knew the correct 
answer from the knowledge of the theoretical $T$-matrix which served as 
input for the evaluation of the observables. Thus it was easy to check 
whether the correct solution was among the ones found numerically, as was 
the case. However, it is worth mentioning that it was not always easy 
to find the proper solution. While for the sets 1, 2, and 4 it was not 
difficult to obtain the correct solution among the various possible ones, 
it was extremely tedious to locate it for set 3. 

In an experimental study, however, the correct solution is not known in 
advance. Therefore, in order to decide, which of the various solutions found 
in general is correct, one has to calculate with the obtained solutions for 
the $T$-matrix 
elements additional observables and compare them to their measured values.
Thus for any complete set one needs to measure 
further observables in order to be able 
to single out a unique solution for the $T$-matrix. For example, for
our first set of Table \ref{solve} it was sufficient to consider one more
observable, namely $x_2^{10}$. But in general it might be necessary, to 
check several additional observables. A further discussion of electromagnetic 
deuteron break-up with regard to complete sets of observables including 
transverse and interference observables will be presented in a forthcoming 
paper.

We hope that this nontrivial example 
demonstrates convincingly the usefulness of our criterion for selecting a 
complete set of observables for a given reaction and that it will provide a 
guideline for polarization studies in the future. 

\section*{Acknowledgment}
Parts of this work were done while H.A. and W.L. were at the
Saskatchewan Accelerator Laboratory and the Department of Physics of the
University of Saskatchewan, Saskatoon 
and while H.A. and E.L.T.  were at the
Dipartimento di Fisica of the Universit\a`a degli Studi di Trento.
In all cases the authors thank the respective institutions
for their hospitality. Furthermore, H.A. and E.L.T. would like to 
thank James Brooke of the Department of Mathematics, 
University of Saskatchewan for useful discussions.

\begin{table}[h]
\caption{Numbering of independent longitudinal $T$-matrix elements 
$(\lambda=0)$ of $d(e,e'N)N$ for the helicity basis.}

\begin{tabular}{ccccccc}
$j$ & $1$ & $2$ & $3$ & $4$ & $5$ & $6$ 
\\
\tableline
&&&&&&\\[-2.ex]
$\lambda_p$ & $\frac{1}{2}$ & $\frac{1}{2}$ & $\frac{1}{2}$ & 
$\frac{1}{2}$ & $\frac{1}{2}$ & $\frac{1}{2}$ 
\\[1.ex] 
$\lambda_n$ & $-\frac{1}{2}$ & $\frac{1}{2}$ & $-\frac{1}{2}$ & 
$\frac{1}{2}$ & $-\frac{1}{2}$ & $\frac{1}{2}$ 
\\[1.ex] 
$\lambda_d$ & $0$ & $0$ & $-1$ & $-1$ & $1$ & $1$ 
\\
\end{tabular}
\label{qnlist}
\end{table}

\begin{table}[h]
\caption{A set of linearly independent longitudinal structure functions
$f_L^{IM} (X)$ of $d(e,e'N)N$ with $IM$-values $(00,11,20,21,22)$ for the 
$A$-type and $(10,11,21,22)$ for the $B$-type observables.}

\begin{tabular}{ccccccccc}
$X$ & $1$ & $xx$ & $xz$ & $y_1$ & $x_1$ & $x_2$ & $z_1$ & $z_2$ 
\\[1.ex]
\tableline
&&&&&&&&\\[-2.ex]
type & $A$ & $A$ & $A$ & $A$ & $B$ & $B$ & $B$ & $B$ 
\\
\end{tabular}
\label{tab1}
\end{table}

\begin{table}[h]
\caption{Complete sets of longitudinal observables of $d(e,e'N)N$ 
by setting $j_0=1$.}

\begin{tabular}{ccccccccccc}
$1^{00}$ & $y_1^{11}$ & $1^{21}$ & $xz^{00}$ & $x_1^{11}$ & $z_1^{11}$ & 
$y_1^{00}$ & $1^{11}$ & $y_1^{21}$ & $x_1^{21}$ & $z_1^{21}$ 
\\[1.ex] 
$1^{20}$ & $xz^{21}$ & $xx^{21}$ & $xz^{20}$ & $x_2^{11}$ & $z_2^{11}$ & 
$y_1^{20}$ & $xx^{11}$ & $xz^{11}$ & $x_2^{21}$ & $z_2^{21}$ 
\\[1.ex]
$xx^{00}$ & $$ & $$ & $$ & $$ & $$ & $$ & $$ & $$ & $$ & $$ 
\\[1.ex]
$xx^{20}$ & $$ & $$ & $$ & $$ & $$ & $$ & $$ & $$ & $$ & $$ 
\\
\end{tabular}
\label{comset}
\end{table}

\begin{table}[h]
\caption{Alternative complete sets of longitudinal observables 
of $d(e,e'N)N$ by setting $j_0=3$.}

\begin{tabular}{ccccccccccc}
$1^{00}$ & $x_1^{11}$ & $1^{21}$ & $x_1^{10}$ & $1^{22}$ & $x_2^{10}$ & 
$y_1^{00}$ & $1^{11}$ & $x_1^{21}$ & $x_1^{22}$ & $z_1^{22}$ 
\\[1.ex] 
$1^{20}$ & $x_2^{11}$ & $z_1^{11}$ & $xz^{00}$ & $xx^{00}$ & $xz^{22}$ & 
$y_1^{20}$ & $z_1^{21}$ & $x_2^{21}$ & $y_1^{22}$ & $z_2^{22}$ 
\\[1.ex]
$z_1^{10}$ & $y_1^{11}$ & $z_2^{11}$ & $xz^{20}$ & $xx^{20}$ & $$ & 
$y_2^{22}$ & $z_2^{21}$ & $y_1^{21}$ & $$ & $$ 
\\[1.ex]
$z_2^{10}$ & $xz^{21}$ & $xx^{21}$ & $$ & $$ & $$ & $$ & $xx^{11}$ & 
$xz^{11}$ & $$ & $$ 
\\[1.ex]
$xx^{22}$ & $$ & $$ & $$ & $$ & $$ & $$ & $$ & $$ & $$ & $$ 
\\
\end{tabular}
\label{comseta}
\end{table}

\begin{table}[h]
\caption{Listing of longitudinal observables of $d(e,e'N)N$ which restrict 
the choice of complete sets. An inclusion of more than eight
of the listed observables into a set of 11 observables would result
automatically in an incomplete set.}

\begin{tabular}{cccccccccccc}
$1^{22}$ & $x_{1}^{10}$ & $x_{2}^{10}$ & $x_{1}^{22}$ & $x_{2}^{22}$ &
$y_1^{22}$ & $z_{1}^{10}$ & $z_{2}^{10}$ & $z_{1}^{22}$ & $z_{2}^{22}$ & 
$xx^{22}$ & $xz^{22}$ \\
\end{tabular}
\label{rstrctset}
\end{table}

\begin{table}[h]
\caption{Selected complete sets for a numerical solution}

\begin{tabular}{ccccccccccc}
$1^{00}$& $1^{11}$ & $1^{21}$ & $y_1^{00}$ & $y_1^{11}$ & $y_1^{21}$ & 
$x_1^{11}$ & $x_1^{21}$ & $z_1^{11}$ & $z_1^{21}$ & $xz^{00}$ 
\\[1.ex]
\tableline
&&&&&&&&&&\\[-2.ex]
$1^{00}$& $1^{11}$ & $1^{21}$ & $1^{22}$ & $y_1^{00}$ & $y_1^{11}$ & 
$y_1^{21}$ & $y_1^{22}$ & $x_1^{10}$ & $x_2^{10}$ & $z_1^{22}$
\\[1.ex]
\hline
&&&&&&&&&&\\[-2.ex]
$1^{00}$& $1^{11}$ & $1^{20}$ & $1^{21}$ & $1^{22}$ & $y_1^{00}$ & 
$y_1^{11}$ & $y_1^{20}$& $y_1^{21}$ & $y_1^{22}$ & $x_1^{10}$ 
\\[1.ex]
\hline
&&&&&&&&&&\\[-2.ex]
$1^{00}$& $1^{11}$ & $1^{20}$ & $1^{21}$ & $1^{22}$ & $y_1^{00}$ & 
$y_1^{11}$ & $y_1^{21}$ & $y_1^{22}$ & $x_1^{10}$ & $z_1^{10}$
\\
\end{tabular}
\label{solve}
\end{table}

\end{document}